\documentclass[a4paper,11pt]{article}
\usepackage{pos}

\title{The Ultraviolet myth}
\author*[a]{Nils-Erik Bomark}
\author[a]{Reidun Renstrøm}

\affiliation[a]{Univeristy of Agder,\\
  Kristansand, Norway}

\emailAdd{nilseb@uia.no}
\abstract{It is very common to introduce quantum physics in an historical context. Though there are advantages to this, it is a problem that many of the stories that have become central to the physics lore are mere pseudo-histories far detached from the real events.

It is about time that we stop uncritically copying these stories and instead make an effort to present the development of quantum physics as it actually was.
This paper deals with one of the most common myths in quantum history, the one about the ultraviolet catastrophe and how it motivated Planck's introduction of quantum physics.

On closer inspection it turns out this story has the time-line completely turned on its head. The ultraviolet catastrophe was first discussed several years after Planck published his radiation law so it played no role in his motivation. Instead Planck was concerned with finding a theoretical derivation of the law for blackbody radiation. This law was first thought to be Wien's radiation law, but when new data disagreed, Planck came up with his own law that fitted the data.

Planck's radiation law first came about as an elaborate fit to data and to derive it he found no other way than to use statistical mechanics and divide the energy that was to be distributed on the atomic oscillators into packages $hf$ so that he could count the number of ways to distribute this energy.
Planck did not consider this a quantization, but merely a mathematical trick to be able to calculate the entropy of the oscillators.

}

\FullConference{The European Physical Society Conference on High Energy Physics (EPS-HEP2023)\\
 21-25 August 2023\\
Hamburg, Germany\\}

\begin{document}
\maketitle

\section{Introduction}

Most physics textbooks state that the origin of quantum physics was a response to crises brought about by grave contradictions between classical physics and experimental fact. However, there were no such crisis that could have motivated Planck in 1900, Einstein in 1905 or Bohr in 1913 to develop their introduction of energy quanta. The contradictions and crisis the textbook version tell and dramatize are constructed; they are plain myths~\cite{Reidun}. 

This paper deals with the myths in the textbook story about Planck's ``birth'' of the quantum physics. 
We also provide a brief historical account of Planck's actual route to his radiation law and the introduction of energy quanta.

\section{The ultraviolet catastrophe and Planck’s hypothesis --- the myth}

The central message of the textbook story regarding the birth of quantum physics can be summarized as follows: By the end of the nineteenth century experimental work had shown how the intensity of the radiation from hot blackbodies varied with wavelength. At the high and low wavelength ends of the spectrum, the intensity decreased rapidly with a maximum in between. In the spring of 1900 Lord Rayleigh used the equipartition principle, a fundamental principle in classical physics,  to derive an equation for the intensity of the different wavelengths $\lambda$,
\begin{equation}
  I(\lambda,T)=c_1  \frac{T}{\lambda^4}
\label{eq:Rayleigh}
\end{equation}
where $c_1$ is a constant and $T$ is the body's temperature.

This law is known as Rayleigh's radiation law. It matched the experimental curves quite well for large wavelengths. However, a significant discrepancy arose for small wavelengths. In Rayleigh's law, the intensity of the radiation approached infinity for small wavelengths, a prediction in Rayleigh’s time called the ultraviolet catastrophe. Even more concerning, the integral of Rayleigh’s equation over all wavelengths yielded an infinite value, indicating an infinitely large total radiated intensity. How devasting for physics.

In an act of desperation Planck found that if he assumed that the electric oscillators that emit the radiation have quantized energy, he could avoid the catastrophe and, even better, derive a correct radiation law. Planck's radical hypothesis, which at that time seemed crazy, brought physics from crisis to success.
 
This story aligns with a certain view of how new theories are develop in physics. It always starts with a crisis brought about by staggering contradiction between experimental facts on the one side and theoretical prediction on the other. Such a crisis motivates physicists to suggest a radical hypothesis to explain the incomprehensible results. By testing the hypothesis' theoretical predictions experimentally, they will either be verified or falsified. If a hypothesis is verified and is the only one that can explain the experimental results, it will be accepted by the physicists and included in the theoretical basis of physics.

\section{The Rayleigh-Jeans-Einstein law --- the real story}
Rayleigh’s proposal was published in a two pages paper in June 1900~\cite{Rayleigh:1900}; {\it Remarks upon the Law of Complete Radiation}. His agenda was to investigate whether the equipartition theorem in statistical mechanics could describe the long wavelengths better than Wien’s law. In the introduction he states about the equipartition principle,
\begin{quote}
  {\it According to this doctrine every mode of vibration should be alike favoured, and although for some reason not yet explained the doctrine fails in general, it seems possible that it may apply to the graver modes.}
\end{quote}
He only claims this might apply to longer wavelengths (which it does!) and he does say the equipartition principle fails in general; this is probably about heat capacity of ideal gases.

He then derives his law and finishes with,
  \begin{quote}
   {\it If we introduce the exponential factor, the complete expression will be,}
  \end{quote}
  \begin{equation}
   I(\lambda,T)=c_1  \frac{T}{\lambda^4}  e^\frac{-c_2}{\lambda T},
  \end{equation}
  where $c_1$ and $c_2$ are constants.

The textbook version of Rayleigh's law lacks the exponential factor! The actual law (above) implied no ``ultraviolet catastrophe'' for small wavelengths. 

In the first part of Einstein's famous 1905 paper~\cite{Einstein:1905cc} where he introduced the light quanta hypothesis, he established that classical physics inexorably led to the law Rayleigh had reached from the equipartition theorem, $c_1  T/\lambda^4$, and Einstein claimed that the spectrum of blackbody radiation could not be explained without a break from classical physics. 

In the same year, Rayleigh published his radiation law without the exponential factor and he now pointed out the difficulty with the formula for high frequencies~\cite{Pais:1982}. Around 1910, after several years with discussions, it was generally accepted that it was not possible to derive a correct radiation law for the high frequencies from classical physics~\cite{Reidun}. The concept ``ultraviolet catastrophe'' was used for the first time in 1911 by Erhenfest~\cite{Reidun} to describe these problems and at the first Solvay Congress in 1911 the reality of energy quanta was discussed. By then most of the physicists realized that energy quanta had changed the course of physics, a decade after Planck's publication.

\section{Planck's route to his radiation law and the energy elements --- the prequel}

In 1860 Gustav R. Kirchhoff stated that one of the main tasks of physics should be to determine the intensity of the different frequencies in the radiation from blackbodies as a function of temperature\cite{Reidun}. In 1896 Wilhelm Wien found a radiation law which was in convincing agreement with observations for all frequencies,
\begin{equation}
  I(\lambda,T)=c_1  \frac{T}{\lambda^5}e^\frac{-c_2}{\lambda T}.
\label{eq:Wien}
\end{equation}

This law worked so well that most physicists believed that the task Kirchhoff had assigned was completed. In the period 1897-1900, Planck published five papers on blackbody radiation, and in the fifth he claimed that he had derived Wien's law. His derivation was based on a relation between the oscillators' entropy and energy without any statistical interpretation. At the beginning of 1900 all seemed well concerning the description of blackbody radiation. 

In October 1900 experimentalists at Berlin's Physikalich-Technische Reichsanstalt reported to Planck that there were deviations from Wien's law for the very low frequencies or large wavelengths~\cite{Pais:1982}. This meant that Wien's law was incorrect and so was his defined relationship between energy and entropy. 

Planck found a new entropy expression for the oscillator mathematically by using the limits of the experimental curve and with this he was able to derive a new law~\cite{Planck:1900a},
\begin{equation}
  I(\lambda,T)=\frac{c_1\lambda^{-5}}{e^{\frac{c_2}{\lambda T}}-1}.
\end{equation}

 This new law agreed perfectly with experiments. At a meeting at the Prussian Academy October 25th 1900, Planck's proposal was presented together with others, among them Rayleigh's~\cite{Reidun}. Rayleigh's law received very little attention. The delegates didn't even notice that Rayleigh's law without the exponential factor fit the data in the low frequency region, where Wien's law failed. 

However, Planck knew that his law was based on an entropy expression that lacked a theoretical foundation, and he felt that he had to derive it at any cost. Once again, he had to derive a perfect radiation law, this time his own. To derive the entropy expression, Planck turned to Boltzmann’s relation between entropy and probability, the law he for so long had ignored, or even tried to disprove. Planck expressed Boltzmann's relation as $S=k\ln{W}$; the entropy, $S$, of a state is given by the probability, $W$, of this state. To calculate the probability, Planck divided the total energy of the oscillators into energy quanta and calculated how many ways those quanta could be distributed~\cite{Planck:1900b}. This procedure led him to his new entropy expression, and his radiation law.

\section{Conclusions}
The idea that physics progress through a series of crisis, is hard to defend. Especially the ultraviolet catastrophe was not a big crisis that motivated Planck to introduce quantum physics.

Planck already had a working radiation law and found a way to derive said law, the first claim that this cannot be done using classical physics came from Einstein in 1905.

%The ultraviolet catastrophe does not play any part in Planck's reasoning, but came later as an argument for Einstein's photons.

%Planck was simply studying blackbody radiation and wanted to find a formula to fit the data and a theoretical motivation for said formula.

\end{document}